\documentstyle[epsfig,12pt]{article}
\newcommand{\Z}{{\sf Z \!\!\! Z}}

\newcommand{\Sign}{\mbox{Sign}}
\newcommand{\PP}{\overline{\Psi} \Psi}
\makeatletter
\@addtoreset{equation}{section}
\makeatother

\title{Meron-Cluster Simulation of a Chiral Phase Transition with Staggered 
Fermions
\footnote{This work is supported in part by funds provided by the U.S.
Department of Energy (D.O.E.) under cooperative research agreements
DE-FC02-94ER40818 and DE-FG02-96ER40945.}}
\author{S. Chandrasekharan$^{{\rm a}}$, J. Cox$^{{\rm b}}$, 
K. Holland$^{{\rm b}}$ and U.-J. Wiese$^{{\rm b}}$ \\ 
\\ 
$^{{\rm a}}$ Department of Physics \\
Duke University \\
Durham, North Carolina 27708, U.S.A. \\ \\
$^{{\rm b}}$
Center for Theoretical Physics, \\
Laboratory for Nuclear Science and Department of Physics \\
Massachusetts Institute of Technology (MIT) \\
Cambridge, Massachusetts 02139, U.S.A. \\ \\
DUKE-TH-99-189, MIT-CTP-2876}
 
\begin{document}
\maketitle
\begin{abstract} \normalsize

We examine a $(3+1)$-dimensional model of staggered lattice fermions with a 
four-fermion interaction and $\Z(2)$ chiral symmetry using the Hamiltonian
formulation. This model cannot be simulated with standard fermion algorithms
because those suffer from a very severe sign problem. We use a new fermion 
simulation technique --- the meron-cluster algorithm --- which solves the sign 
problem and leads to high-precision numerical data. We investigate the finite 
temperature chiral phase transition and verify that it is in the universality 
class of the 3-d Ising model using finite-size scaling.

\end{abstract}
 
\maketitle
 
\newpage

\section{Introduction}

The numerical simulation of lattice fermions is a notoriously difficult problem
which is the major stumbling block in solving QCD and other fermionic field 
theories. The standard method is to integrate out the fermions and to simulate 
the resulting bosonic problem with a non-local action. In several cases of 
physical interest --- for example, for QCD with an odd number of flavors or 
with non-zero chemical potential --- the bosonic Boltzmann factor may become 
negative or even complex and thus cannot be interpreted as a probability. When 
the sign or the complex phase of the Boltzmann factor is included in measured 
observables, the numerical simulation suffers from severe cancellations 
resulting in a sign problem. The standard fermion algorithms are incapable of 
exploring such models. As a consequence, QCD is usually simulated with an even 
number of flavors and at zero chemical potential. Even in the absence of a sign
problem, the simulation of fermions is difficult. For example, lattice QCD
simulations suffer from critical slowing down when one approaches the chiral 
limit in which the quarks become massless. In particular, this makes it 
difficult to identify the universality class of the finite temperature QCD 
chiral phase transition. 

Even in simpler models with four-fermion interactions the identification of the
finite temperature critical behavior is a non-trivial issue \cite{Kog98}. A 
model with $N$ fermion flavors shows mean-field behavior in the $N = \infty$ 
limit. On the other hand, at finite $N$ one finds the non-trivial critical 
behavior that one expects based on dimensional reduction and standard 
universality arguments. For example, in \cite{Kog99} it has been verified that 
the chiral phase transition in a $(2+1)$-d four-fermion model with $N = 4$ and 
$\Z(2)$ chiral symmetry is in the universality class of the 2-d Ising model. 
Due to the fermion sign problem, standard fermion simulation methods often do 
not work in models with a too small number of flavors.

In this paper we apply a recently developed technique \cite{Cha99} for solving 
the sign problem to a $(3+1)$-d model of staggered fermions using the 
Hamiltonian formulation. The model has $N = 2$ flavors and a $\Z(2)$ chiral 
symmetry that is spontaneously broken at low temperatures. The fermion 
determinant can be negative in this model. Hence, due to the sign problem 
standard fermion algorithms fail in this case. Our algorithm is the only 
numerical method available to simulate this model. In this method we do not 
integrate out the fermions but describe them in a Fock state basis. The 
resulting bosonic model of fermion occupation numbers interacts locally, but 
has a non-local fermion permutation sign resulting from the Pauli exclusion 
principle. Standard numerical methods would suffer from severe cancellations of
positive and negative contributions to the partition function. Like other 
cluster methods, our algorithm decomposes a configuration of fermion occupation
numbers into clusters which can be flipped independently. Under a cluster flip 
an occupied site becomes empty and vice versa. The main idea of the 
meron-cluster algorithm is to construct the clusters such that they affect the 
fermion sign independent of each other when they are flipped. In addition, it 
must always be possible to flip the clusters into a reference configuration 
with a positive sign. A cluster whose flip changes the fermion sign is referred
to as a meron because it can be viewed as a half-instanton. If a configuration 
contains a meron-cluster, its contribution to the partition function is 
canceled by the contribution of the configuration that one obtains when the 
meron-cluster is flipped. The observables that we consider get non-zero 
contributions from the zero- and two-meron sectors only. Our algorithm ensures 
that configurations with more than two merons are never generated, which leads 
to an exponential gain in statistics and to a complete solution of the sign 
problem. 

Like other cluster algorithms the meron algorithm substantially reduces 
critical slowing down. This allows us to work directly in the chiral limit. As 
a result, we can study the nature of the chiral phase transition in great 
detail. The $\Z(2)$ chiral symmetry is spontaneously broken at low temperatures
and gets restored in the high-temperature phase. As expected, the system close 
to the finite temperature critical point is in the universality class of the 
3-d Ising model. We verify this in a high-precision finite-size scaling 
investigation of the chiral susceptibility.

The paper is organized as follows. In section 2 we introduce the staggered
fermion Hamiltonian, derive the path integral representation of the partition 
function, and discuss the fermion sign as well as relevant observables. Section
3 contains the description of the meron-cluster algorithm and the corresponding
improved estimators. In section 4, we present the results of numerical 
simulations and extract the critical behavior from the finite-size scaling of 
the chiral susceptibility. Finally, section 5 contains our conclusions.

\section{The Staggered Fermion Model}

Let us consider staggered fermions hopping on a 3-d cubic spatial lattice with 
$V = L^3$ sites $x$ ($L$ even) and with periodic or antiperiodic spatial 
boundary conditions. We start in the Hamiltonian formulation and then derive a 
path integral on a $(3+1)$-d Euclidean space-time lattice. The fermions are 
described by creation and annihilation operators $\Psi_x^+$ and $\Psi_x$ with 
standard anticommutation 
relations
\begin{equation}
\{\Psi_x^+,\Psi_y^+\} = \{\Psi_x,\Psi_y\} = 0, \, 
\{\Psi_x^+,\Psi_y\} = \delta_{xy}.
\end{equation}
The staggered fermion Hamilton operator takes the form 
\begin{equation}
H = \sum_{x,i} h_{x,i} + m \sum_x (-1)^{x_1+x_2+x_3} \Psi_x^+ \Psi_x,
\end{equation}
that is a sum of nearest-neighbor couplings $h_{x,i}$ and a mass term $m \PP$. 
In the following we work directly in the chiral limit, $m = 0$, and only use 
$\PP$ as an observable. The term $h_{x,i}$ couples the fermion operators at the
lattice sites $x$ and $x+\hat i$, where $\hat i$ is a unit-vector in the 
$i$-direction, and
\begin{equation}
h_{x,i} = \frac{1}{2} \eta_{x,i}(\Psi_x^+ \Psi_{x+\hat i} + 
\Psi_{x+\hat i}^+ \Psi_x) + G (\Psi_x^+ \Psi_x - \frac{1}{2})
(\Psi_{x+\hat i}^+ \Psi_{x+\hat i} - \frac{1}{2}).
\end{equation}
Here $\eta_{x,1} = 1$, $\eta_{x,2} = (-1)^{x_1}$ and $\eta_{x,3} = 
(-1)^{x_1 + x_2}$ are the standard staggered fermion sign factors, and $G$ is a
four-fermion coupling constant. The system has a conserved fermion number
\begin{equation}
N = \sum_x \Psi_x^+ \Psi_x,
\end{equation}
because $[H,N] = 0$. Besides the $U(1)$ fermion number symmetry, the model has
a $\Z(2)$ chiral symmetry, which (up to a phase) simply shifts $\Psi_x^+$ and 
$\Psi_x$ by one lattice spacing in all three directions. This changes the sign 
of $\PP$ but leaves the $m=0$ Hamiltonian invariant. There are also other 
$\Z(2)$ symmetries which correspond to discrete flavor transformations. For a 
detailed discussion of the symmetries of staggered fermions in the Hamiltonian 
formulation we refer to \cite{Sus77}.

To construct a path integral for the partition function, we decompose the 
Hamilton operator into six terms $H = H_1 + H_2 + ... + H_6$, 
with
\begin{equation}
H_i = \!\!\! \sum_{\stackrel{x = (x_1,x_2,x_3)}{x_i even}} \!\!\! h_{x,i}, 
\, \,
H_{i+3} = \!\!\! \sum_{\stackrel{x = (x_1,x_2,x_3)}{x_i odd}} \!\!\! h_{x,i}.
\end{equation}
The individual contributions to a given $H_i$ commute with each other, but two 
different $H_i$ do not commute. Using the Suzuki-Trotter formula we express the
fermionic partition function at inverse temperature $\beta$ as
\begin{equation}
Z_f = \mbox{Tr} [\exp(- \beta H)] = \lim_{M \rightarrow \infty} \mbox{Tr} 
[\exp(- \epsilon H_1) \exp(- \epsilon H_2) ... \exp(- \epsilon H_6)]^M.
\end{equation}
We have introduced $6M$ Euclidean time slices with $\epsilon = \beta/M$ being 
the lattice spacing in the Euclidean time direction. Following Jordan and 
Wigner \cite{Jor28} we represent the fermion operators by Pauli matrices
\begin{equation}
\Psi_x^+ = \sigma_1^3 \sigma_2^3 ... \sigma_{l-1}^3 \sigma_l^+, \,
\Psi_x = \sigma_1^3 \sigma_2^3 ... \sigma_{l-1}^3 \sigma_l^-, \,
n_x = \Psi_x^+ \Psi_x = \frac{1}{2}(\sigma_l^3 + 1),
\end{equation}
with
\begin{equation}
\sigma_l^\pm = \frac{1}{2} (\sigma_l^1 \pm i \sigma_l^2), \, 
[\sigma_l^i,\sigma_m^j] = 2 i \delta_{lm} \epsilon_{ijk} \sigma_l^k.
\end{equation}
Here $l$ labels the lattice point $x$. The Jordan-Wigner representation
requires an ordering of the lattice points. For example, one can label the 
point $x = (x_1,x_2,x_3)$ (with $x_i = 0,1,...,L-1$) by $l = 1 + x_1 + x_2 L + 
x_3 L^2$. It should be pointed out that the Jordan-Wigner representation works 
in any dimension. In one dimension the lattice points are, of course, naturally
ordered, but even in higher dimensions the physics is completely independent of
the arbitrary ordering. We now insert complete sets of fermion Fock states 
between the factors $\exp(- \epsilon H_i)$. Each site is either empty or 
occupied, i.e. $n_x$ has eigenvalue $0$ or $1$. In the Pauli matrix 
representation this corresponds to eigenstates $|0\rangle$ and $|1\rangle$ of 
$\sigma_l^3$ with $\sigma_l^3 |0\rangle = - |0\rangle$ and $\sigma_l^3 
|1\rangle = |1\rangle$. The transfer matrix is a product of factors
\begin{equation}
\label{transfer}
\exp(- \epsilon h_{x,i}) = \exp(\frac{\epsilon G}{4})
\left(\begin{array}{cccc}
\exp(- \frac{\epsilon G}{2}) & 0 & 0 & 0 \\ 
0 & \cosh \frac{\epsilon}{2} & \Sigma \sinh \frac{\epsilon}{2} & 0 \\ 
0 & \Sigma \sinh \frac{\epsilon}{2} & \cosh \frac{\epsilon}{2} & 0 \\ 
0 & 0 & 0 & \exp(- \frac{\epsilon G}{2}) \end{array} \right),
\nonumber \\ \
\end{equation}
which is a $4 \times 4$ matrix in the Fock space basis $|00\rangle$,
$|01\rangle$, $|10\rangle$ and $|11\rangle$ of two sites $x$ and $x+\hat i$.
Here $\Sigma = \eta_{x,i} \sigma_{l+1}^3 \sigma_{l+2}^3 ... \sigma_{m-1}^3$ 
includes the local sign $\eta_{x,i}$ as well as a non-local string of Pauli 
matrices running over consecutive labels between $l$ and $m$, where $l$ labels 
the lattice point $x$ and $m$ labels $x+\hat i$. Note that $\Sigma$ is diagonal
in the occupation number basis.

The partition function is now expressed as a path integral
\begin{equation}
Z_f = \sum_n \Sign[n] \exp(- S[n]),
\end{equation}
over configurations of occupation numbers $n(x,t) = 0,1$ on a $(3+1)$-d 
space-time lattice of points $(x,t)$. The Boltzmann factor
\begin{eqnarray}
\exp(- S[n])&=&\!\!\!\!\! \prod_{\stackrel{x = (x_1,x_2,x_3)}
{x_1 even, t = 6m}} \!\!\!\!\!\!\!\!
\exp\{- s[n(x,t),n(x+\hat 1,t),n(x,t+1),n(x+\hat 1,t+1)]\} \nonumber \\
&\times&\!\!\!\!\! \prod_{\stackrel{x = (x_1,x_2,x_3)}
{x_2 even, t = 6m+1}} \!\!\!\!\!\!\!\!
\exp\{- s[n(x,t),n(x+\hat 2,t),n(x,t+1),n(x+\hat 2,t+1)]\} \nonumber \\
&\times&\!\!\!\!\! \prod_{\stackrel{x = (x_1,x_2,x_3)}
{x_3 even, t = 6m+2}} \!\!\!\!\!\!\!\!
\exp\{- s[n(x,t),n(x+\hat 3,t),n(x,t+1),n(x+\hat 3,t+1)]\} \nonumber \\
&\times&\!\!\!\!\! \prod_{\stackrel{x = (x_1,x_2,x_3)}
{x_1 odd, t = 6m+3)}} \!\!\!\!\!\!\!\!
\exp\{- s[n(x,t),n(x+\hat 1,t),n(x,t+1),n(x+\hat 1,t+1)]\} \nonumber \\
&\times&\!\!\!\!\! \prod_{\stackrel{x = (x_1,x_2,x_3)}
{x_2 odd, t = 6m+4}} \!\!\!\!\!\!\!\!
\exp\{- s[n(x,t),n(x+\hat 2,t),n(x,t+1),n(x+\hat 2,t+1)]\} \nonumber \\
&\times&\!\!\!\!\! \prod_{\stackrel{x = (x_1,x_2,x_3)}
{x_3 odd, t = 6m+5}} \!\!\!\!\!\!\!\!
\exp\{- s[n(x,t),n(x+\hat 3,t),n(x,t+1),n(x+\hat 3,t+1)]\}, \nonumber \\ \
\end{eqnarray}
(with $m = 0,1,...,M-1$) is a product of space-time plaquette contributions 
with
\begin{eqnarray}
\label{Boltzmann}
&&\exp(- s[0,0,0,0]) = \exp(- s[1,1,1,1]) = \exp(- \frac{\epsilon G}{2}),
\nonumber \\
&&\exp(- s[0,1,0,1]) = \exp(- s[1,0,1,0]) = \cosh \frac{\epsilon}{2},
\nonumber \\
&&\exp(- s[0,1,1,0]) = \exp(- s[1,0,0,1]) = \sinh \frac{\epsilon}{2}.
\end{eqnarray}
All the other Boltzmann weights are zero, which implies several constraints on
allowed configurations. Note that here we have dropped the trivial overall 
factor $\exp(\epsilon G/4)$ that appeared in eq.(\ref{transfer}).

The sign of a configuration, $\Sign[n]$, also is a product of space-time 
plaquette contributions
$\mbox{sign}[n(x,t),n(x+\hat i,t),n(x,t+1),n(x+\hat i,t+1)]$ with
\begin{eqnarray}
&&\mbox{sign}[0,0,0,0]) = \mbox{sign}[0,1,0,1]) = \mbox{sign}[1,0,1,0]) = 
\mbox{sign}[1,1,1,1]) = 1, \nonumber \\
&&\mbox{sign}[0,1,1,0]) = \mbox{sign}[1,0,0,1]) = \Sigma.
\end{eqnarray}
It should be noted that $\Sigma$ gets contributions from all lattice points 
with labels between $l$ and $m$. This seems to make an evaluation of the 
fermion sign rather tedious. Also, it is not a priori obvious that $\Sign[n]$ 
is independent of the arbitrarily chosen order of the lattice points. 
Fortunately, there is a simple way to compute $\Sign[n]$, which is directly 
related to the Pauli exclusion principle and which is manifestly 
order-independent. In fact, $\Sign[n]$ has a topological meaning. The occupied 
lattice sites define fermion world-lines which are closed around the Euclidean 
time direction. Of course, during their Euclidean time evolution fermions can 
interchange their positions, and the fermion world-lines define a permutation 
of particles. The Pauli exclusion principle dictates that the fermion sign is 
just the sign of that permutation. If we work with antiperiodic spatial 
boundary conditions, $\Sign[n]$ receives an extra minus-sign for every fermion 
world-line that crosses a spatial boundary. Figure 1 shows two configurations 
of fermion occupation numbers in $(1+1)$ dimensions. The first configuration 
corresponds to two fermions at rest and has $\Sign[n] = 1$. In the second 
configuration two fermions interchange their positions with one fermion 
stepping over the spatial boundary. If one uses periodic spatial boundary 
conditions this configuration has $\Sign[n] = - 1$. Note that the same 
configuration would have $\Sign[n] = 1$ when antiperiodic boundary conditions 
are used.
\begin{figure}[ht]
\hbox{
\hspace{2.7cm}
${\rm Sign[n]}=1$
\hspace{5.0cm}
${\rm Sign[n]}=-1$
}  
\begin{center}
\hbox{
\epsfig{file=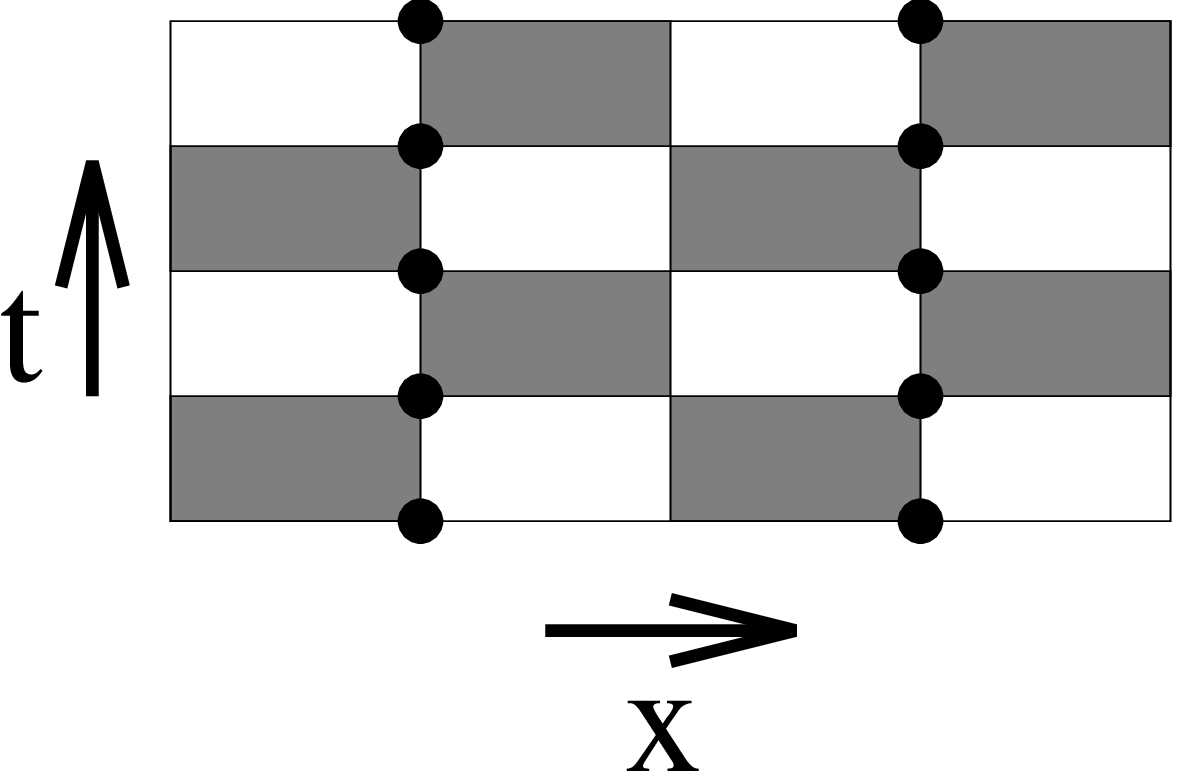,
width=4.5cm,angle=270,
bbllx=0,bblly=0,bburx=225,bbury=337}
\hspace{1.0cm}
\epsfig{file=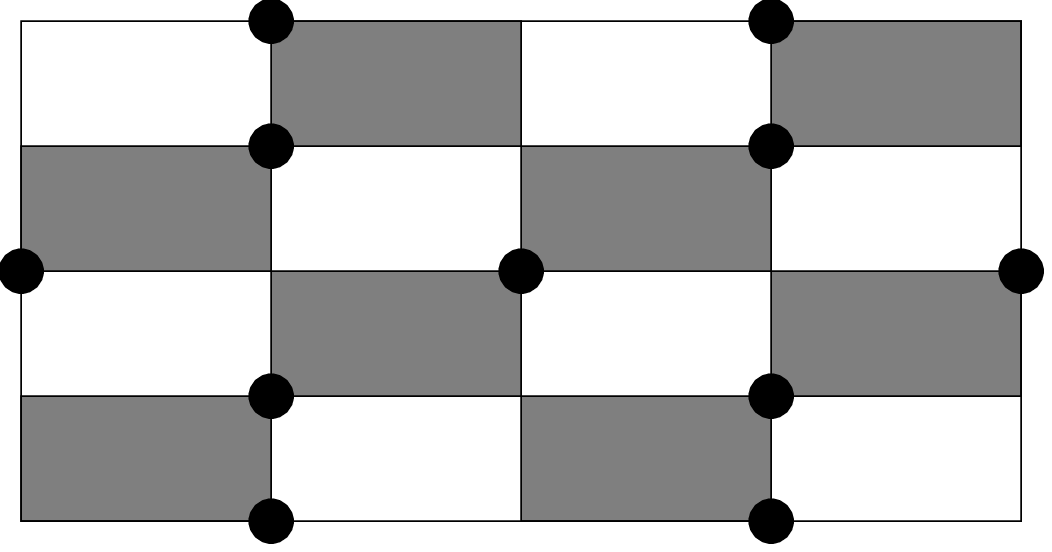,
width=4.5cm,angle=270,
bbllx=0,bblly=0,bburx=225,bbury=337}
}  
\end{center}
\caption{\it Two configurations of fermion occupation numbers in $(1+1)$ 
dimensions. The shaded plaquettes carry the interaction. The dots represent 
occupied sites. In the second configuration two fermions interchange their 
positions. With periodic spatial boundary conditions this configuration has
$\Sign[n] = - 1$.}
\end{figure}

The expectation value of a fermionic observable $O[n]$ is given by
\begin{equation}
\langle O \rangle_f = \frac{1}{Z_f} \sum_n O[n] \Sign[n] \exp(- S[n]).
\end{equation}
Quantities of physical interest are the chiral condensate
\begin{equation}
\PP[n] = \frac{\epsilon}{6} \sum_{x,t} (-1)^{x_1 + x_2 + x_3} 
(n(x,t) - \frac{1}{2}),
\end{equation}
and the corresponding chiral susceptibility
\begin{equation}
\chi = \frac{1}{\beta V} \langle (\overline{\Psi} \Psi)^2 \rangle_f.
\end{equation}

Up to now we have derived a path integral representation for the fermion
system in terms of bosonic occupation numbers and a fermion sign factor that
encodes Fermi statistics. The system without the sign factor is bosonic and
is characterized by the positive Boltzmann factor $\exp(- S[n])$. Here the 
bosonic model is a quantum spin system with the Hamiltonian
\begin{equation}
H = \sum_{x,i} (S_x^1 S_{x+\hat i}^1 + S_x^2 S_{x+\hat i}^2 + 
G S_x^3 S_{x+\hat i}^3), 
\end{equation}
where $S_x^i = \frac{1}{2} \sigma_l^i$ is a spin $1/2$ operator associated with
the lattice site $x$ that was labeled by $l$. The case $G = 1$ corresponds to
the isotropic antiferromagnetic quantum Heisenberg model, $G = 0$ represents 
the quantum XY-model, and $G = - 1$ corresponds to an isotropic ferromagnet.
In the language of the spin model, the chiral condensate turns into the
staggered magnetization
\begin{equation}
\PP = \frac{\epsilon}{6} \sum_{x,t} (-1)^{x_1+x_2+x_3} S_x^3.
\end{equation}
For sufficiently large $G$, for example in an antiferromagnet with $G = 1$, 
the staggered magnetization gets a non-zero expectation value at sufficiently 
low temperature, thus breaking the bosonic analog of chiral symmetry. It will 
turn out that the fermion sign does not change this behavior, and indeed chiral
symmetry is spontaneously broken in the fermionic model as well.

\section{The Meron-Cluster Algorithm}

Let us first discuss the nature of the fermion sign problem. The fermionic path
integral $Z_f = \sum_n \Sign[n] \exp(-S[n])$ includes the fermion sign factor 
$\Sign[n] = \pm 1$ as well as a positive Boltzmann factor $\exp(-S[n])$ that 
contains the action $S[n]$ of the corresponding bosonic model with partition 
function $Z_b = \sum_n \exp(-S[n])$. A fermionic observable $O[n]$ is obtained 
in a simulation of the bosonic ensemble as
\begin{equation}
\langle O \rangle_f = \frac{1}{Z_f} \sum_n O[n] \Sign[n] \exp(-S[n]) =
\frac{\langle O \ \Sign \rangle}{\langle \Sign \rangle}.
\end{equation}
The average sign in the simulated bosonic ensemble is given by
\begin{equation}
\langle \Sign \rangle = \frac{1}{Z_b}{\sum_n \Sign[n] \exp(-S[n])} = 
\frac{Z_f}{Z_b} = \exp(- \beta V \Delta f).
\end{equation}
The last equality points to the heart of the sign problem. The expectation 
value of the sign is exponentially small in both the volume $V$ and the inverse
temperature $\beta$ because the difference between the free energy densities 
$\Delta f = f_f - f_b$ of the fermionic and bosonic systems is necessarily 
positive.

Even in an ideal simulation of the bosonic ensemble which generates $N$ 
completely uncorrelated configurations, the relative statistical error of the 
sign is
\begin{equation}
\frac{\Delta \Sign}{\langle \Sign \rangle} = 
\frac{\sqrt{\langle \Sign^2 \rangle - \langle \Sign \rangle^2}}
{\sqrt{N} {\langle \Sign \rangle}} = \frac{\exp(\beta V \Delta f)}{\sqrt{N}}.
\end{equation}
Here we have used $\Sign^2 = 1$. To determine the average sign with sufficient 
accuracy one needs to generate on the order of $N = \exp(2 \beta V \Delta f)$ 
configurations. For large volumes and small temperatures this is impossible in
practice. It is possible to solve one half of the problem if one can match any 
contribution $-1$ with another contribution $1$ to give $0$, such that only a 
few unmatched contributions $1$ remain. Then effectively $\Sign = 0,1$ and 
hence $\Sign^2 = \Sign$. This reduces the relative error to
\begin{equation}
\frac{\Delta \Sign}{\langle \Sign \rangle} = 
\frac{\sqrt{\langle \Sign \rangle - \langle \Sign \rangle^2}}
{\sqrt{N'} {\langle \Sign \rangle}} =
\frac{\exp(\beta V \Delta f/2)}{\sqrt{N'}}.
\end{equation}
One gains an exponential factor in statistics, but one still needs to generate 
$N' = \sqrt{N} = \exp(\beta V \Delta f)$ independent configurations in order to
accurately determine the average sign.\footnote{The fact that an improved 
estimator alone cannot solve the sign problem was pointed out to one of the 
authors by H. G. Evertz a long time ago.} This is because one generates 
exponentially many vanishing contributions before one encounters a contribution
$1$. As explained below, in our cluster algorithm an explicit matching of 
contributions $-1$ and $1$ is achieved using an improved estimator. This solves
one half of the sign problem. In a second step involving a Metropolis decision,
our algorithm ensures that contributions $0$ and $1$ occur with similar 
probabilities. This saves another exponential factor in statistics and solves 
the other half of the sign problem.

The meron-cluster fermion algorithm is based on a cluster algorithm for the 
corresponding bosonic model without the sign factor. Bosonic quantum spin 
systems can be simulated very efficiently with cluster algorithms 
\cite{Wie92,Eve93,Eve97}. The first cluster algorithm for lattice fermions was 
described in \cite{Wie93}. These algorithms can be implemented directly in the 
Euclidean time continuum \cite{Bea96}, i.e. the Suzuki-Trotter discretization 
is not even necessary. The same is true for the meron-cluster algorithm. Here 
we discuss the algorithm for discrete time. The idea behind the algorithm is to
decompose a configuration into clusters which can be flipped independently. 
Each lattice site belongs to exactly one cluster. When the cluster is flipped, 
the occupation number of all the sites on the cluster is changed from $n(x,t)$ 
to $1- n(x,t)$, i.e. a cluster flip turns occupied into empty sites and vice 
versa. The decomposition of the lattice into clusters results from connecting 
neighboring sites on each individual space-time interaction plaquette following
probabilistic cluster rules. A sequence of connected sites defines a cluster. 
In this case the clusters are sets of closed loops. The cluster rules are 
constructed to obey detailed balance. To show this we first write the plaquette
Boltzmann factors as
\begin{eqnarray}
\label{cluster}
&&\!\!\!\exp(- s[n(x,t),n(x+\hat i,t),n(x,t+1),n(x+\hat i,t+1)]) = \nonumber \\
&&\!\!\!A \delta_{n(x,t),n(x,t+1)} \delta_{n(x+\hat i,t),n(x+\hat i,t+1)} +
B \delta_{n(x,t),1-n(x+\hat i,t)} \delta_{n(x,t+1),1-n(x+\hat i,t+1)} +
\nonumber \\
&&\!\!\!C \delta_{n(x,t),n(x,t+1)} \delta_{n(x+\hat i,t),n(x+\hat i,t+1)}
\delta_{n(x,t),1-n(x+\hat i,t)} + 
D \delta_{n(x,t),n(x+\hat i,t+1)} \delta_{n(x+\hat i,t),n(x,t+1)} +
\nonumber \\
&&\!\!\!E \delta_{n(x,t),n(x,t+1)} \delta_{n(x+\hat i,t),n(x+\hat i,t+1)}
\delta_{n(x,t),n(x+\hat i,t)}.
\end{eqnarray}
The various $\delta$-functions specify which sites are connected and thus
belong to the same cluster. The quantities $A,B,...,E$ determine the relative 
probabilities for different cluster break-ups of an interaction plaquette. We
only allow break-ups which generate legal configurations under cluster flips.
For example, $A$ determines the probability with which sites are connected with
their time-like neighbors, while $B$ and $D$ determine the probabilities for 
connections with space-like or diagonal neighbors, respectively. The quantities
$C$ and $E$ determine the probabilities to put all four sites of a plaquette 
into the same cluster. This is possible for plaquette configurations 
$[0,1,0,1]$ or $[1,0,1,0]$ with a probability proportional to $C$ and for 
configurations $[0,0,0,0]$ or $[1,1,1,1]$ with a probability proportional to 
$E$. The cluster rules are illustrated in table 1.
\begin{table}[ht]
% space before first and after last column: 1.5pc
% space between columns: 3.0pc (twice the above)
%\setlength{\tabcolsep}{1.5pc}
% -----------------------------------------------------
% adapted from TeX book, p. 241
\newlength{\digitwidth} \settowidth{\digitwidth}{\rm 0}
\catcode`?=\active \def?{\kern\digitwidth}
% -----------------------------------------------------
  \begin{center}
    \leavevmode
%    \tiny
    \begin{tabular}{|c|c|c|}
	\hline
	weight&configuration&break-ups\\
	\hline\hline
	$\exp\left(-\frac{\epsilon G}{2}\right)$
	&\hspace{0.4cm}\begin{minipage}[c]{2.5cm}
		\epsfig{file=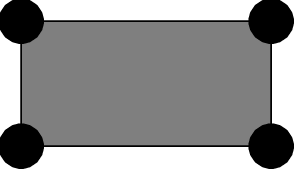,width=1.5cm,angle=270,
		bbllx=-5,bblly=0,bburx=55,bbury=85}
	\end{minipage}
	&\hspace{0.4cm}\begin{minipage}[c]{2.5cm}
		\vbox{\begin{center}
		\epsfig{file=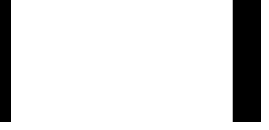,width=1.5cm,angle=270,
		bbllx=-20,bblly=0,bburx=55,bbury=85} 

		\hspace{-0.2cm}A
		\end{center}}
	\end{minipage}
	\hspace{0.4cm}\begin{minipage}[c]{2.5cm}
		\vbox{\begin{center}
		\epsfig{file=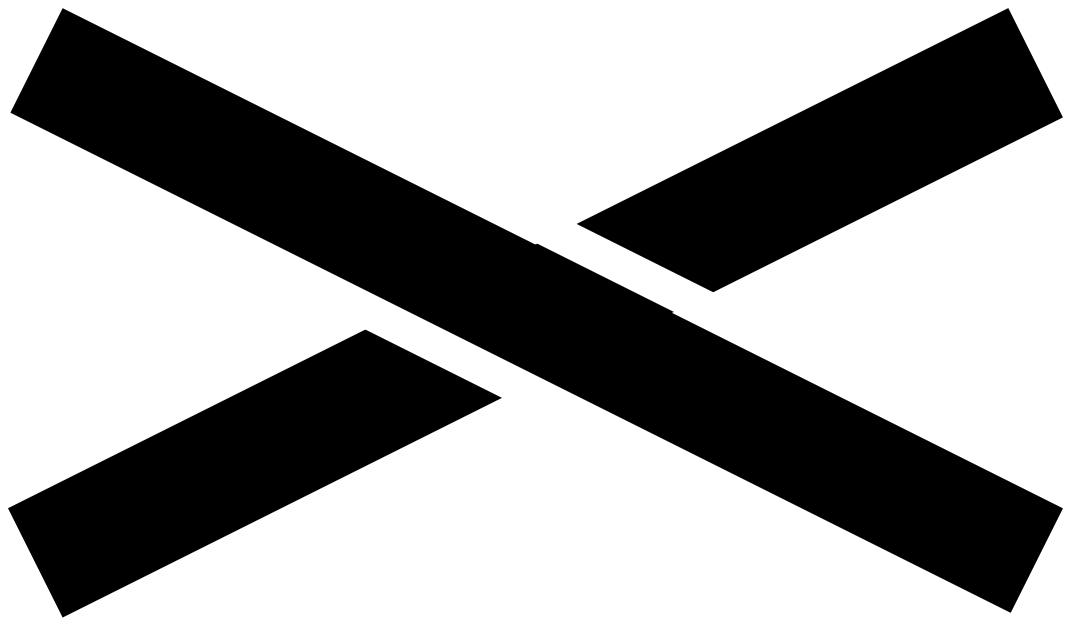,width=1.5cm,angle=270,
		bbllx=-50,bblly=50,bburx=250,bbury=300} 

		\hspace{-0.2cm}D
		\end{center}}
	\end{minipage}
	\hspace{0.4cm}\begin{minipage}[c]{2.5cm}
		\vbox{\begin{center}
		\epsfig{file=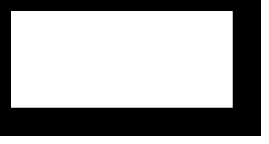,width=1.5cm,angle=270,
		bbllx=-20,bblly=0,bburx=55,bbury=85} 

		\hspace{-0.2cm}E
		\end{center}}
	\end{minipage}
	\\
	\hline
	$\cosh\left(\frac{\epsilon}{2}\right)$
	&\hspace{0.4cm}\begin{minipage}[c]{2.5cm}
		\epsfig{file=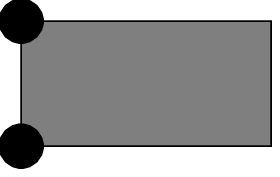,width=1.5cm,angle=270,
		bbllx=-5,bblly=0,bburx=55,bbury=85}
	\end{minipage}
	&\hspace{0.4cm}\begin{minipage}[c]{2.5cm}
		\vbox{\begin{center}
		\epsfig{file=table_1_2.eps,width=1.5cm,angle=270,
		bbllx=-20,bblly=0,bburx=55,bbury=85} 

		\hspace{-0.2cm}A
		\end{center}}
	\end{minipage}
	\hspace{0.4cm}\begin{minipage}[c]{2.5cm}
		\vbox{\begin{center}
		\epsfig{file=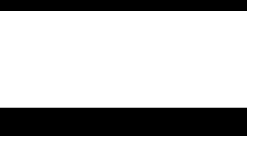,width=1.5cm,angle=270,
		bbllx=-20,bblly=0,bburx=55,bbury=85} 

		\hspace{-0.2cm}B
		\end{center}}
	\end{minipage}
	\hspace{0.4cm}\begin{minipage}[c]{2.5cm}
		\vbox{\begin{center}
		\epsfig{file=table_1_4.eps,width=1.5cm,angle=270,
		bbllx=-20,bblly=0,bburx=55,bbury=85} 

		\hspace{-0.2cm}C
		\end{center}}
	\end{minipage}
	\\
	\hline
	$\sinh\left(\frac{\epsilon}{2}\right)$
	&\hspace{0.4cm}\begin{minipage}[c]{2.5cm}
		\epsfig{file=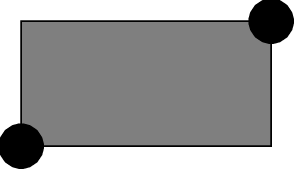,width=1.5cm,angle=270,
		bbllx=-5,bblly=0,bburx=55,bbury=85}
	\end{minipage}
	&\hspace{0.4cm}\begin{minipage}[c]{2.5cm}
		\vbox{\begin{center}
		\epsfig{file=table_3_2.eps,width=1.5cm,angle=270,
		bbllx=-20,bblly=0,bburx=55,bbury=85} 

		\hspace{-0.2cm}B
		\end{center}}
	\end{minipage}
	\hspace{0.4cm}\begin{minipage}[c]{2.5cm}
		\vbox{\begin{center}
		\epsfig{file=table_1_3.eps,width=1.5cm,angle=270,
		bbllx=-50,bblly=50,bburx=250,bbury=300} 

		\hspace{-0.2cm}D
		\end{center}}
	\end{minipage}
	\\
	\hline   
    \end{tabular}
\caption{\it Cluster break-ups of various plaquette configurations together 
with their relative probabilities $A,B,...,E$. The dots represent occupied 
sites and the fat lines are the cluster connections.}
\end{center}
\end{table}

Inserting the expressions from eq.(\ref{Boltzmann}) one finds
\begin{eqnarray}
\label{balance}
&&\exp(- s[0,0,0,0]) = \exp(- s[1,1,1,1]) = \exp(- \frac{\epsilon G}{2}) = 
A + D + E, \nonumber \\
&&\exp(- s[0,1,0,1]) = \exp(- s[1,0,1,0]) = \cosh \frac{\epsilon}{2} = 
A + B + C, \nonumber \\
&&\exp(- s[0,1,1,0]) = \exp(- s[1,0,0,1]) = \sinh \frac{\epsilon}{2} = B + D.
\end{eqnarray}
For example, the probability to connect the sites with their time-like 
neighbors on a plaquette with configuration $[0,0,0,0]$ or $[1,1,1,1]$ is
$A/(A+D+E)$, while the probability for a connection with their diagonal
neighbor is $D/(A+D+E)$. All sites on such a plaquette are put into the same
cluster with probability $E/(A+D+E)$. Similarly, the probability for connecting
space-like neighbors on a plaquette with configuration $[0,1,1,0]$ or 
$[1,0,0,1]$ is $B/(B+D)$ and the probability for diagonal connections is
$D/(B+D)$. 

Eq.(\ref{cluster}) can be viewed as a representation of the original model as a
random cluster model. The cluster algorithm operates in two steps. First, a 
cluster break-up is chosen for each space-time interaction plaquette according 
to the above probabilities. This effectively replaces the original Boltzmann 
weight of a plaquette configuration with a set of constraints represented by 
the $\delta$-functions associated with the chosen break-up. The constraints 
imply that occupation numbers of connected sites can only be changed together. 
In the second step of the algorithm every cluster is flipped with probability 
$1/2$. When a cluster is flipped the occupation numbers of all sites that 
belong to the cluster are changed. Eq.(\ref{balance}) ensures that the cluster 
algorithm obeys detailed balance. To determine $A,B,...,E$ we distinguish three
cases. For $G \geq 1$ we solve eq.(\ref{balance}) by
\begin{equation}
\label{solution1}
A = \exp(- \frac{\epsilon G}{2}), \ B = \sinh \frac{\epsilon}{2}, \
C = \exp(- \frac{\epsilon}{2}) - \exp(- \frac{\epsilon G}{2}), \ D = E = 0.
\end{equation}
For $- 1 \leq G \leq 1$ we use
\begin{eqnarray}
\label{solution2}
&&A = \frac{1}{2}[\exp(- \frac{\epsilon G}{2}) + \exp(- \frac{\epsilon}{2})],
\ B = \frac{1}{2}[\exp(\frac{\epsilon}{2}) - \exp(- \frac{\epsilon G}{2})], \ 
C = 0, \nonumber \\
&&D = \frac{1}{2}[\exp(- \frac{\epsilon G}{2}) - \exp(- \frac{\epsilon}{2})], \
E = 0, \end{eqnarray}
and, finally, for $G \leq - 1$
\begin{equation}
\label{solution3}
A = \cosh \frac{\epsilon}{2}, \ B = C = 0, \ D = \sinh \frac{\epsilon}{2}, \
E = \exp(- \frac{\epsilon G}{2}) - \exp(\frac{\epsilon}{2}).
\end{equation}
As an example, let us consider the antiferromagnetic quantum Heisenberg model,
i.e. $G = 1$, and hence
\begin{equation}
A = \exp(- \frac{\epsilon}{2}), \ B = \sinh \frac{\epsilon}{2}, \ 
C = D = E = 0.
\end{equation}
Consequently, on plaquette configurations $[0,0,0,0]$ or $[1,1,1,1]$ one
always chooses time-like connections between sites, and for configurations
$[0,1,1,0]$ or $[1,0,0,1]$ one always chooses space-like connections. For
configurations $[0,1,0,1]$ or $[1,0,1,0]$ one chooses time-like connections
with probability $p = A/(A+B) = 2/[1 + \exp(\epsilon/2)]$ and space-like 
connections with probability $1-p = B/(A+B)$. Indeed, this is the algorithm 
that was used in \cite{Wie94}. It is extremely efficient, has almost no 
detectable autocorrelations, and its dynamical exponent for critical slowing 
down is compatible with zero.

Let us now consider the effect of a cluster flip on the fermion sign. It is
obvious that the flip of a cluster either leads to a sign change or it leaves 
the sign unchanged. In general, the effect of the flip of a specific cluster
on the fermion sign depends on the orientation of the other clusters. For
example, a cluster whose flip does not change the sign now, may very well
change the sign after other clusters have been flipped. In other words, the
clusters affect each other in their effect on the fermion sign. This makes it
very difficult to understand the effect of the various cluster flips on the 
topology of the fermion world-lines and thus on $\Sign[n]$. As a consequence, 
for the most general model described above, we don't know how to solve the
fermion sign problem. However, there are clusters whose effect on $\Sign[n]$ is
independent of the orientation of all the other clusters. Specifically, these 
are those clusters that do not contain any diagonal break-ups. To ensure that
no clusters contain such break-ups we limit ourselves to models which have
$D = 0$. According to eq.(\ref{solution1}) this is the case for $G \geq 1$. 

Once we forbid diagonal cluster break-ups, i.e. when $D=0$, the clusters have
a remarkable property with far reaching consequences: each cluster can then be 
characterized by its effect on the fermion sign independent of the orientation 
of all the other clusters. We refer to clusters whose flip changes $\Sign[n]$
as merons, while clusters whose flip leaves $\Sign[n]$ unchanged are called 
non-merons. The flip of a meron-cluster permutes the fermions and changes the
topology of the fermion world-lines. The term meron has been used before to
denote half-instantons. For example, the meron-clusters in the algorithm for
the 2-d $O(3)$ model at non-zero vacuum angle $\theta$ are indeed 
half-instantons \cite{Bie95}. A configuration with an odd permutation of 
fermions has $\Sign[n] = - 1$ and can be viewed as an instanton, while a
configuration with an even permutation of fermions is topologically trivial and
has $\Sign[n] = 1$. The flip of a meron-cluster changes an instanton into a
topologically trivial configuration and therefore is a half-instanton. In
$(2+1)$ dimensions particles can have any statistics. The anyon statistics is
characterized by a phase factor $\exp(i \theta H)$, where $H$ is the 
integer-valued Hopf number. In that case, the flip of a meron-cluster changes
the Hopf number by one, and the meron is indeed a half-Hopf-instanton. The 
number of merons in a configuration is always even. This follows when one 
considers flipping all clusters, and thus changing the occupation of all 
lattice sites, which leaves the fermion sign unchanged. If the number of merons
would be odd, flipping all clusters would change the fermion sign, which 
implies that the number of merons must be even. An example of a meron-cluster 
is given in figure 2, which contains the same fermion configurations as figure 
1. When the meron-cluster is flipped the first configuration with 
$\Sign[n] = 1$ turns into the second configuration with $\Sign[n] = - 1$.
\begin{figure}[ht]
\hbox{
\hspace{2.7cm}
${\rm Sign[n]}=1$
\hspace{5.0cm}
${\rm Sign[n]}=-1$
}  
\begin{center}
\hbox{
\epsfig{file=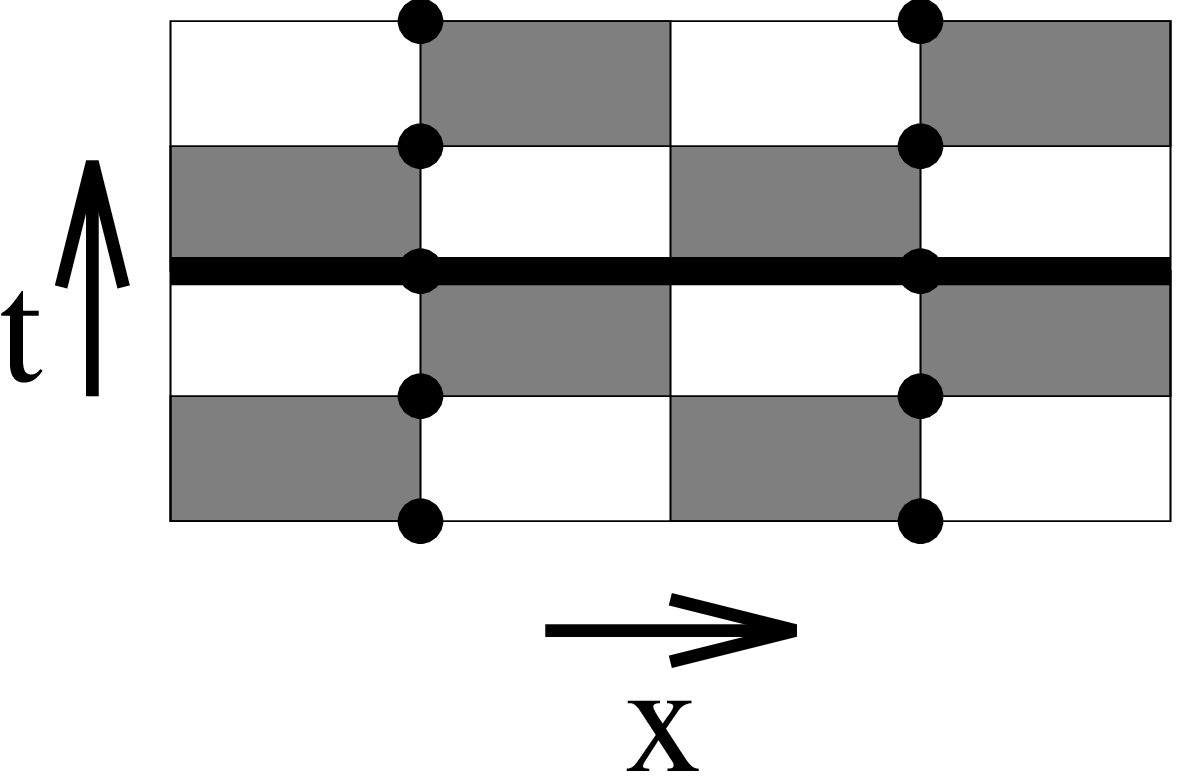,
width=4.5cm,angle=270,
bbllx=0,bblly=0,bburx=225,bbury=337}
\hspace{1.0cm}
\epsfig{file=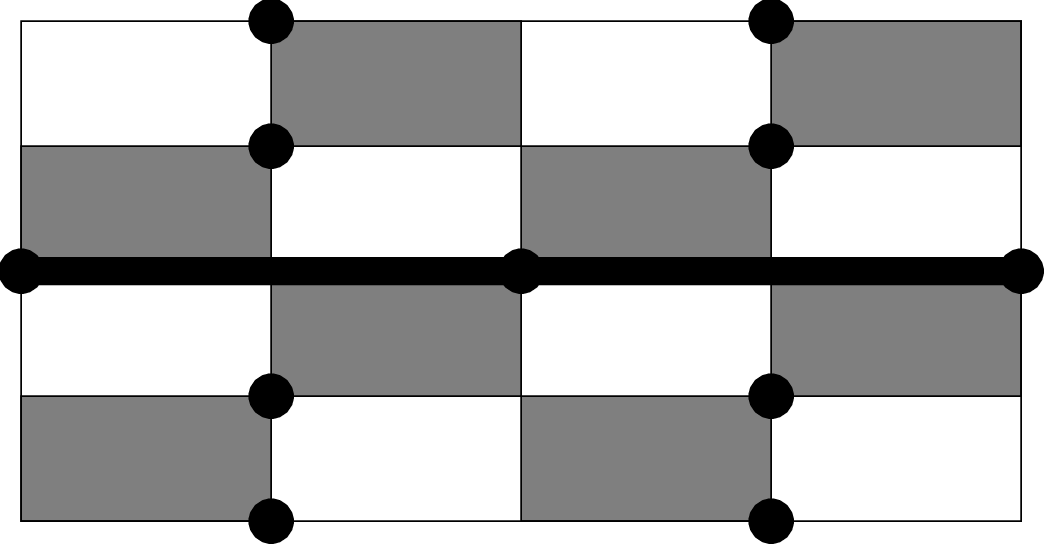,
width=4.5cm,angle=270,
bbllx=0,bblly=0,bburx=225,bbury=337}
}  
\end{center}
\caption{\it The same fermion configurations as in figure 1 together with a 
meron-cluster represented by the fat line. The other clusters are not shown.
Flipping the meron-cluster changes one configuration into the other and changes
the fermion sign.}
\end{figure}

The meron concept alone allows us to gain an exponential factor in statistics.
Since all clusters can be flipped independently, one can construct an improved 
estimator for $\Sign[n]$ by averaging analytically over the $2^{N_C}$ 
configurations obtained by flipping the $N_C$ clusters in the configuration in 
all possible ways. For configurations that contain merons the average 
$\Sign[n]$ is zero because flipping a single meron leads to a cancellation of
contributions $\pm 1$. Hence, only the configurations without merons contribute
to $\Sign[n]$. The vast majority of configurations contains merons and now 
contributes an exact $0$ to $\Sign[n]$ instead of a statistical average of 
contributions $\pm 1$. In this way the improved estimator leads to an 
exponential gain in statistics.

Still, as it stands it is not guaranteed that the contributions from the
zero-meron sector will always be positive. With no merons in the configuration
it is clear that the fermion sign remains unchanged under cluster flip, but one
could still have $\Sign[n] = - 1$. Fortunately, there is a way to guarantee 
that $\Sign[n] = 1$ in the zero-meron sector. In that case the contributions to
$\Sign[n]$ are $0$ from configurations containing meron-clusters and $1$ from 
configurations without merons. According to the previous discussion, this 
solves one half of the fermion sign problem. Let us describe how one can make 
sure that a configuration without merons always has $\Sign[n] = 1$. This is 
possible when one does not allow the cluster break-ups characterized by the 
amplitudes $D$ and $E$ in eq.(\ref{cluster}). This again limits us to models 
with $G \geq 1$, for which we have $D = E = 0$ according to 
eq.(\ref{solution1}). The remaining cluster break-ups with amplitudes $A, B, C$
have a very important property. They guarantee that sites inside a cluster obey
a pattern of staggered occupation, i.e. either the even sites (with $x_1 + x_2 
+ x_3$ even) within the cluster are all occupied and the odd sites are all 
empty, or the even sites are all empty and the odd sites are all occupied. This
guarantees that the clusters can be flipped such that one reaches the totally 
staggered reference configuration in which at all times all even sites are 
occupied and all odd sites are empty. In the corresponding antiferromagnet 
this configuration represents a completely ordered state with a staggered 
magnetization. In the half-occupied reference configuration (which is the first
configuration in figure 1) all fermions are at rest, no fermions are permuted 
during the Euclidean time evolution, and thus $\Sign[n] = 1$. Since any 
configuration can be turned into the reference configuration by appropriate
cluster flips, this is particularly true for configurations without merons. 
Since the totally staggered configuration has $\Sign[n] = 1$ and the fermion 
sign remains unchanged when a non-meron-cluster is flipped, all configurations 
without merons have $\Sign[n] = 1$. Instead of a sequence of $\pm 1$ for 
$\Sign[n]$, we now have contributions $0$ and $1$. As discussed before, this 
only solves one half of the fermion sign problem. Before we can solve the other
half of the problem we must discuss improved estimators for the physical 
observables.

Let us consider an improved estimator for $(\PP[n])^2 \Sign[n]$ which is needed
to determine the chiral susceptibility $\chi$. The total chiral condensate, 
$\PP[n] = \sum_C \PP_C$, is a sum of cluster contributions
\begin{equation} 
\PP_C = \frac{\epsilon}{6} \sum_{(x,t) \in C} (-1)^{x_1+x_2+x_3}
(n(x,t) - \frac{1}{2}).
\end{equation}
When a cluster is flipped, its condensate contribution changes sign. In a 
configuration without merons, where $\Sign[n] = 1$ for all relative cluster 
flips, the average of $(\PP[n])^2 \Sign[n]$ over all $2^{N_C}$ configurations 
is $\sum_C |\PP_C|^2$. For configurations with two merons the average is 
$2 |\PP_{C_1}||\PP_{C_2}|$ where $C_1$ and $C_2$ are the two meron-clusters. 
Configurations with more than two merons do not contribute to $(\PP[n])^2 
\Sign[n]$. The improved estimator for the susceptibility is hence given by
\begin{equation}
\label{chi}
\chi = \frac{\langle \sum_C |\PP_C|^2 \delta_{N,0} + 2 |\PP_{C_1}||\PP_{C_2}| 
\delta_{N,2} \rangle}{V \beta \langle \delta_{N,0} \rangle},
\end{equation}
where $N$ is the number of meron-clusters in a configuration. Thus, to 
determine $\chi$ one must only sample the zero- and two-meron sectors. 

The probability to find a configuration without merons is exponentially small 
in the space-time volume since it is equal to $\langle \Sign \rangle$. Thus, 
although we have increased the statistics tremendously with the improved 
estimators, without a second step one would still need an exponentially large 
statistics to accurately determine $\chi$. Fortunately, the numerator in
equation (\ref{chi}) receives contributions from the zero- and two-meron 
sectors only, while the denominator gets contributions only from the zero-meron
sector. One can hence restrict oneself to the zero- and two-meron sectors and 
never generate configurations with more than two merons. This enhances both the
numerator and the denominator by a factor that is exponentially large in the
volume, but leaves the ratio of the two invariant. One purpose of the second
step of the meron-cluster algorithm is to eliminate all configurations with 
more than two merons. To achieve this, we start with an initial configuration 
with zero or two merons. For example, a completely occupied configuration has 
no merons. We then visit all plaquette interactions one after the other and 
choose new pair connections between the four sites according to the above 
cluster rules. If the new connection increases the number of merons beyond two,
it is not accepted and the old connection is kept for that plaquette. This 
procedure obeys detailed balance because configurations with more than two 
merons do not contribute to the observable we consider. This simple reject step
eliminates almost all configurations with weight $0$ and is the essential step 
to solve the other half of the fermion sign problem. 

Assuming a dilute gas of meron and non-meron-clusters of typical space-time 
volume $|C|$ one finds a ratio $p(0)/p(2) \propto (|C|/V \beta)^2$ of the 
probabilities $p(0)$ and $p(2)$ to have zero or two merons. Hence, as long as
the cluster size does not grow with the space-time volume, most configurations 
would have two merons and therefore would still have weight $0$. Without 
further improvements, one would still need statistics quadratic (but no longer 
exponential) in the space-time volume to get an accurate average sign. The 
remaining problem can be solved with a reweighting technique similar to the 
one used in \cite{Bie95}. To enhance the zero-meron configurations in a 
controlled way, we introduce trial probabilities $p_t(0)$ and $p_t(2)$ which 
determine the relative weight of the zero- and two-meron sector. The trial 
distribution $p_t(N)$ for $N>2$ is set to infinity. The distribution $p_t(N)$ 
is used in a Metropolis accept-reject step for the newly proposed pair 
connection on a specific plaquette interaction. A new pair connection that 
changes the meron number from $N$ to $N'$ is accepted with probability 
$p = \mbox{min}[1,p_t(N)/p_t(N')]$. In particular, configurations with $N'>2$ 
are never generated because then $p_t(N') = \infty$ and $p=0$. After visiting 
all plaquette interactions, each cluster is flipped with probability $1/2$ 
which completes one update sweep. After reweighting, the zero- and two-meron 
configurations appear with similar probabilities. This completes the second 
step in our solution of the fermion sign problem. The reweighting of the zero- 
and two-meron configurations is taken into account in the final expression for 
the chiral susceptibility as
\begin{equation}
\chi = \frac{\langle \sum_C |\PP_C|^2 \delta_{N,0} \ p_t(0) + 
2 |\PP_{C_1}||\PP_{C_2}| \delta_{N,2} \ p_t(2) \rangle}
{V \beta \langle \delta_{N,0} \ p_t(0) \rangle}.
\end{equation}
The optimal ratio $p_t(0)/p_t(2)$, which minimizes the statistical error of
$\chi$, can be estimated by gradually increasing the volume and the inverse 
temperature to their desired values.

\section{Numerical Results}

We have simulated the staggered fermion model with $G = 1$ on antiperiodic
spatial volumes $L^3$ with $L = 4,6,...,16$ and at various inverse temperatures
$\beta \in [0.5,1.2]$ which includes the critical point. In the Euclidean time
direction we have used $M = 4$, i.e. 24 time-slices. In all cases, we have 
performed at least 1000 thermalization sweeps followed by 10000 measurements.
One sweep consists of a new choice of the cluster connections on each 
interaction plaquette and a flip of each cluster with probability $1/2$. To 
estimate an optimal ratio $p_t(0)/p_t(2)$ of the reweighting probabilities,
we have first performed runs without reweighting. The observed relative 
probabilities of the zero- and two-meron sectors were used as an estimate for 
$p_t(0)/p_t(2)$ in production runs. Especially for the larger volumes, 
reweighting is necessary in order to obtain accurate results. 

Some of our data for the chiral susceptibility $\chi$ are contained in table 2.
The table also includes the non-reweighted $\langle \Sign \rangle$, the value 
of the used reweighting factor $p_t(0)/p_t(2)$ as well as the reweighted 
$\langle \Sign \rangle_r$.
\begin{table}
\begin{center}
\begin{tabular}{|c|c|c|c|c|c|}
\hline
$L$ & $\beta$ & $\langle \Sign \rangle$ & $p_t(0)/p_t(2)$ & 
$\langle \Sign \rangle_r$ & $\chi$ \\
\hline
\hline
  4 & 0.6 & 0.838(1) & 0.5/0.5 & 0.845(1) & 0.554(1) \\
\hline
  4 & 0.7 & 0.710(3) & 0.5/0.5 & 0.726(2) & 0.936(2) \\
\hline
  4 & 0.8 & 0.534(4) & 0.5/0.5 & 0.566(2) & 1.678(5) \\
\hline
  4 & 0.9 & 0.357(3) & 0.5/0.5 & 0.405(2) & 3.13(1) \\
\hline
  4 & 0.948 & 0.282(2) & 0.3/0.7 & 0.537(3) & 4.22(2) \\
\hline
  6 & 0.948 & 0.0556(7) & 0.2/0.8 & 0.398(3) & 9.83(8) \\
\hline
  8 & 0.948 & 0.0020(4) & 0.1/0.9 & 0.361(8) & 16.6(5) \\
\hline
 10 & 0.948 & --- & 0.1/0.9 & 0.178(3) & 26.5(8) \\
\hline
 12 & 0.948 & --- & 0.05/0.95 & 0.17(1) & 37(1) \\
\hline
 14 & 0.948 & --- & 0.02/0.98 & 0.20(1) & 51(1) \\
\hline
 16 & 0.948 & --- & 0.01/0.99 & 0.22(2) & 65(2) \\
\hline
\end{tabular}
\end{center}
\caption{\it Numerical results for the non-reweighted $\langle \Sign \rangle$, 
the reweighted $\langle \Sign \rangle_r$ and $\chi$ obtained with a reweighting
factor $p_t(0)/p_t(2)$ on lattices of spatial size $L$ at inverse temperature 
$\beta$. For the larger volumes the non-reweighted $\langle \Sign \rangle$ is 
too small to be measured.}
\end{table}
Note that $\langle \Sign \rangle_r$ is the fraction of zero-meron 
configurations that the algorithm generates by sampling the zero- and two-meron
sectors only. This quantity is typically a lot bigger than the original 
non-reweighted $\langle \Sign \rangle$, which is the fraction of zero-meron 
configurations in the space of all configurations including those with many 
merons. In particular, in large space-time volumes the staggered fermion model 
suffers from a very severe sign problem. 
\begin{figure}[ht]
\vbox{
\begin{center}
\psfig{figure=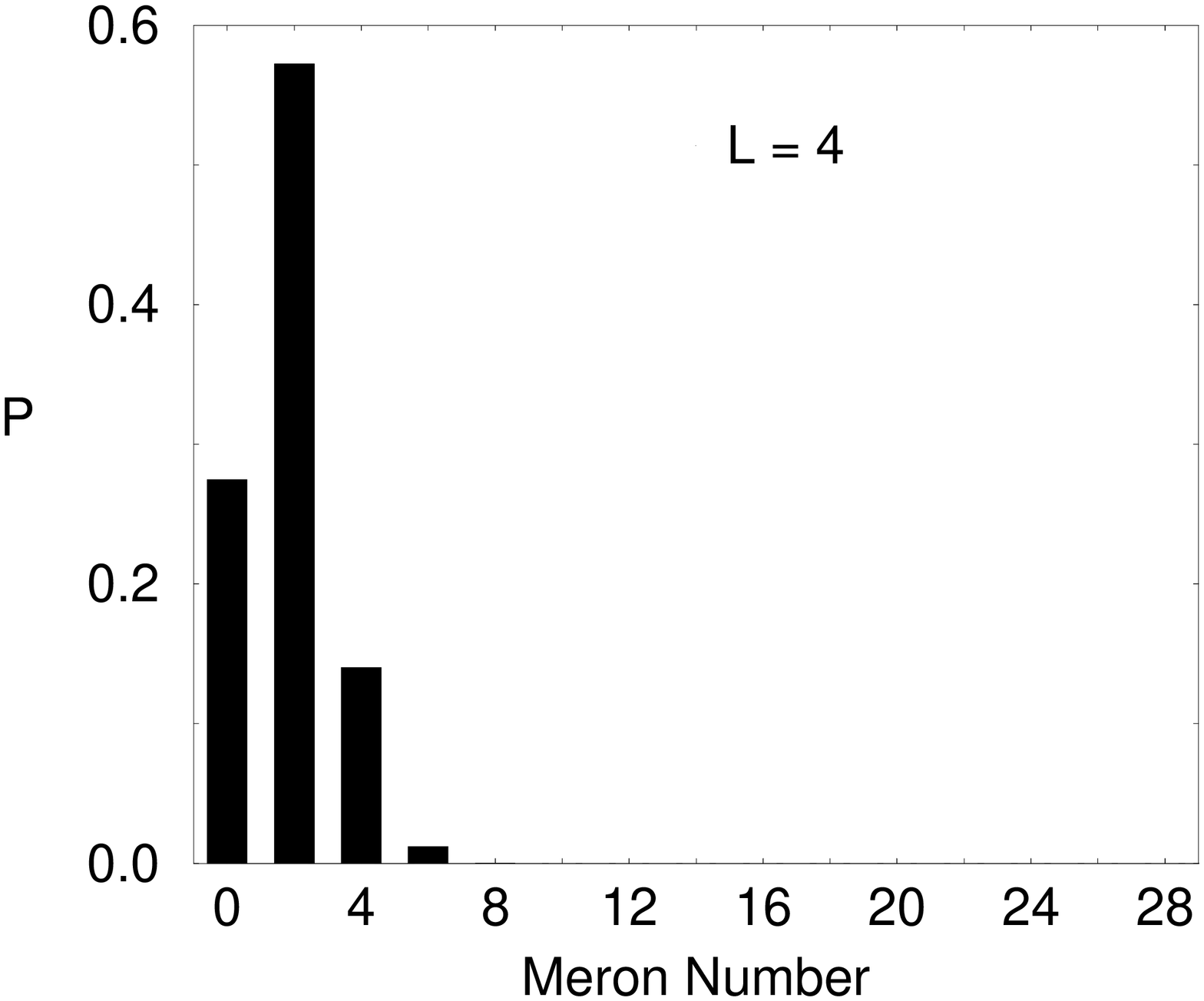,height=2.5in,width=2in,angle=270}
\psfig{figure=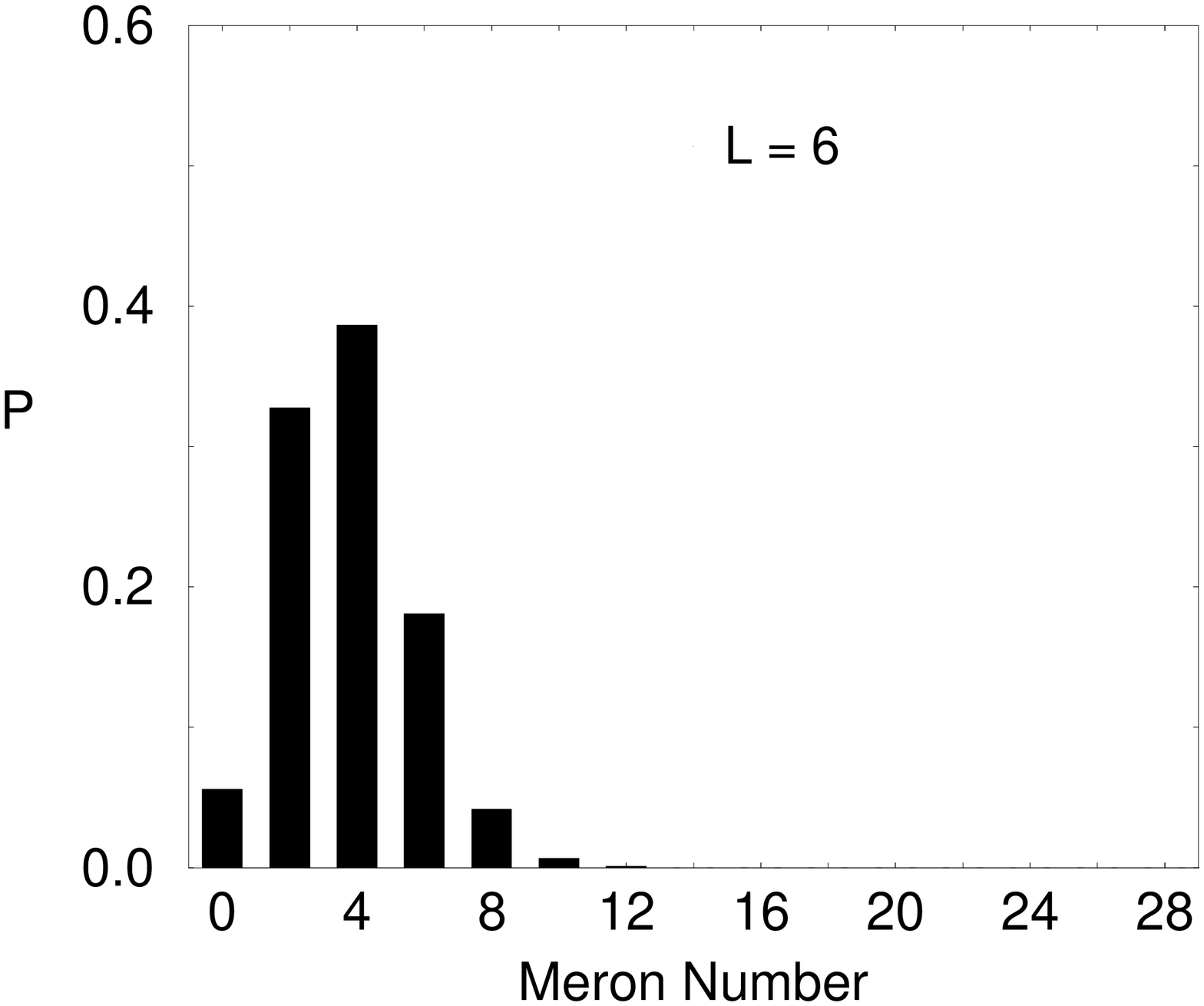,height=2.5in,width=2in,angle=270}
\psfig{figure=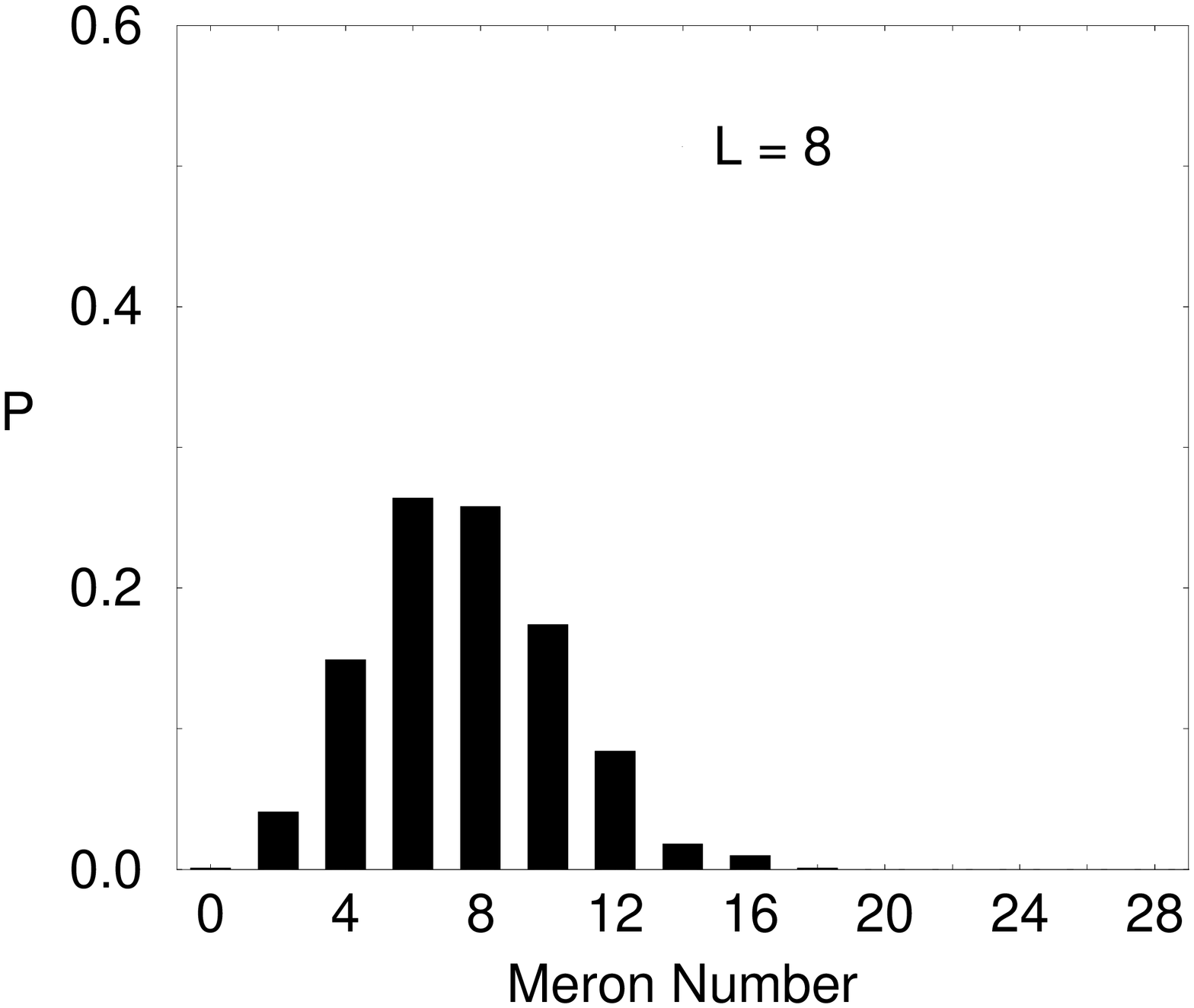,height=2.5in,width=2in,angle=270}
\psfig{figure=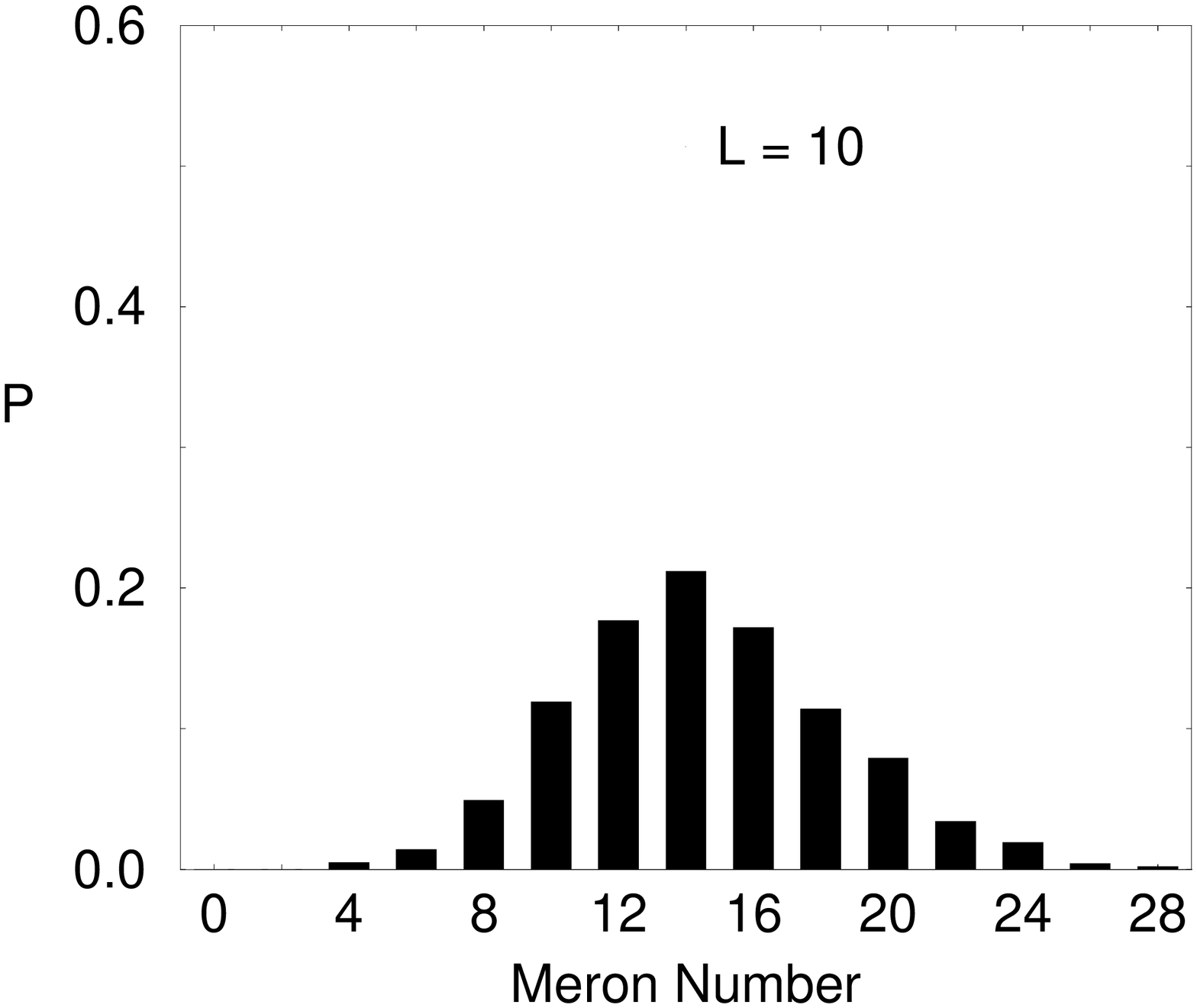,height=2.5in,width=2in,angle=270}
\caption{\it The probability of having a certain number of merons for various
spatial sizes $L = 4,6,8$ and $10$ at $\beta = 0.948$.}
\end{center}
}
\end{figure}
Figure 3 shows the probability to have a certain number of merons in an 
algorithm that samples all meron-sectors without reweighting. For small volumes
the zero-meron sector and hence $\langle \Sign \rangle$ are relatively large, 
while multi-meron configurations are rare. On the other hand, in larger volumes
the vast majority of configurations has a large number of merons and hence 
$\langle \Sign \rangle$ is exponentially small. For example, an extrapolation 
from smaller volumes gives a rough estimate for the non-reweighted 
$\langle \Sign \rangle \approx 10^{-20}$ on the $16^3$ lattice at 
$\beta = 0.948$, while the reweighted $\langle \Sign \rangle_r = 0.22(2)$. 
Hence, to achieve a similar accuracy without the meron-cluster algorithm one 
would have to increase the statistics by a factor $10^{40}$, which is obviously
impossible in practice. In fact, at present there is no other method that can 
be used to simulate this model.

Figure 4 shows the chiral susceptibility $\chi$ as a function of $\beta$ for 
various spatial sizes $L$. At high temperatures (small $\beta$) $\chi$ is 
almost independent of the volume, indicating that chiral symmetry is unbroken. 
On the other hand, at low temperatures $\chi$ increases with the volume, which 
implies that chiral symmetry is spontaneously broken. 
\begin{figure}[ht]
\begin{center}
\psfig{figure=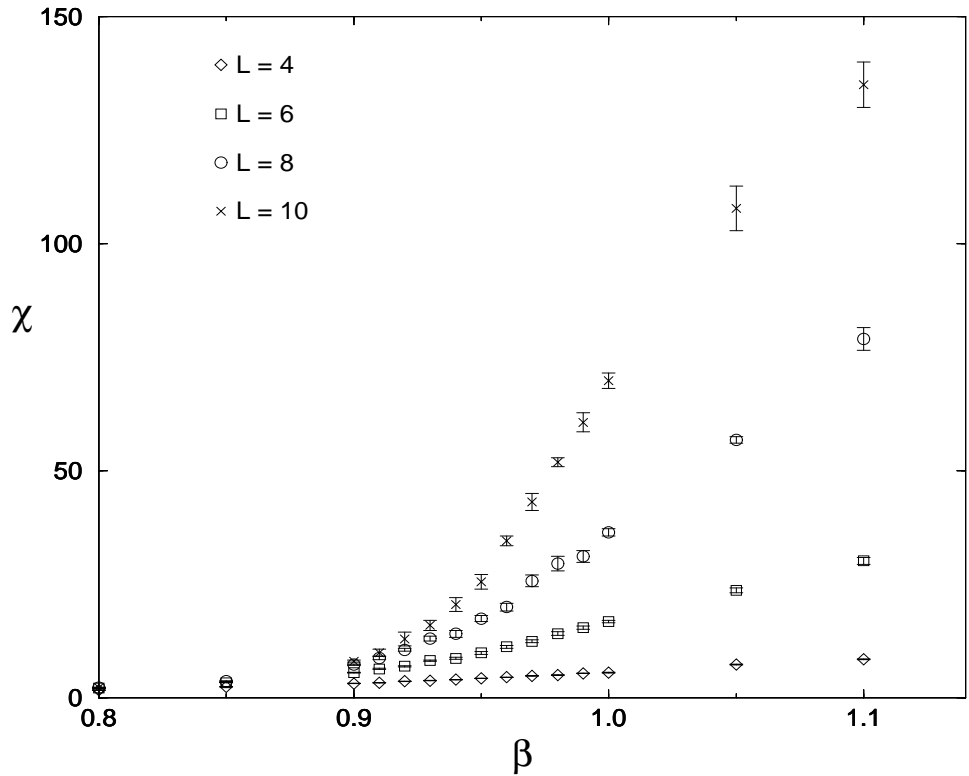,height=5in,width=4in,angle=270}
\caption{\it The chiral susceptibility $\chi$ as a function of the inverse 
temperature $\beta$ for various spatial sizes $L = 4,6,8$ and $10$.}
\end{center}
\end{figure}
To study the critical behavior in detail, we have performed a finite-size
scaling analysis for $\chi$ focusing on a narrow range $\beta \in [0.9,0.98]$
around the critical point. Since a $\Z(2)$ chiral symmetry gets spontaneously 
broken at finite temperature in this $(3+1)$-d model, one expects to find the 
critical behavior of the 3-d Ising model. The corresponding finite-size scaling
formula valid close to $\beta_c$ is \cite{Blo95}
\begin{eqnarray}
\label{scaling}
&&\chi(L,\beta) = a(x) + b(y) L^{\gamma/\nu}, 
\nonumber \\
&&a(x) = a_0 + a_1 x + a_2 x^2 + ..., \ x = \beta - \beta_c,
\nonumber \\
&&b(y) = b_0 + b_1 y + b_2 y^2 + ..., \ y = (\beta - \beta_c) L^{1/\nu}.
\end{eqnarray}
For the 3-d Ising model the critical exponents are given by $\nu = 0.630(1)$ 
and $\gamma/\nu = 1.963(3)$ \cite{Blo95}. Fitting our data, we find 
$\nu = 0.63(4)$ and $\gamma/\nu = 1.98(2)$, which indicates that the chiral 
transition of the staggered fermion model is indeed in the 3-d Ising 
universality class. The fit gives $\beta_c = 0.948(3)$. In figure 5 we have 
taken the values of the critical exponents from the 3-d Ising model, and we 
have plotted $\chi/L^{\gamma/\nu}$ as a function of $y = (\beta - \beta_c) 
L^{1/\nu}$. For large enough $L$ one can neglect the term $a(x)$ in 
eq.(\ref{scaling}) and one obtains $\chi/L^{\gamma/\nu} = b(y)$. We have varied
the value of $\beta_c$ and found that indeed all data can be collapsed on one 
universal curve. The resulting critical inverse temperature is 
$\beta_c = 0.948(3)$ in agreement with the previous fit. 
\begin{figure}[ht]
\begin{center}
\psfig{figure=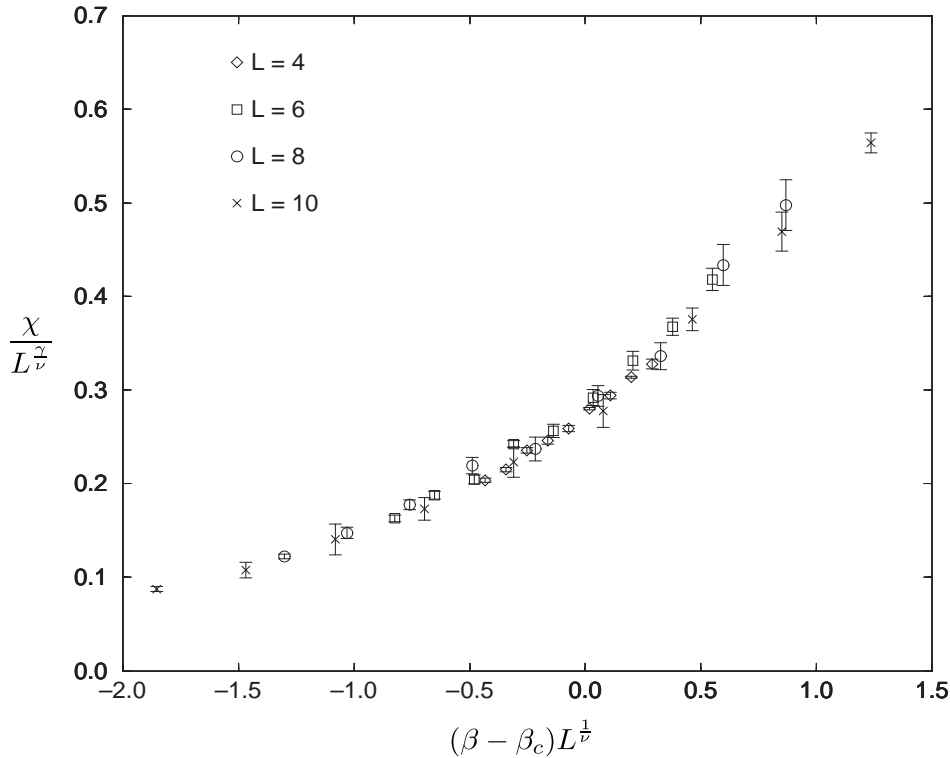,height=4in,width=5in,angle=0}
\caption{\it Finite-size scaling behavior of the chiral susceptibility $\chi$. 
The data for various spatial sizes $L = 4,6,8$ and $10$ fall on one universal 
curve.}
\end{center}
\end{figure}
At the estimated value of $\beta_c$, we have performed
simulations on larger spatial volumes up to $16^3$. The results for $\chi$ are
shown in figure 6 together with a fit that gives an independent estimate of 
$\gamma/\nu = 1.98(2)$. All of this supports the claim that the chiral 
transition in our model is Ising-like.
\begin{figure}[ht]
\begin{center}
\psfig{figure=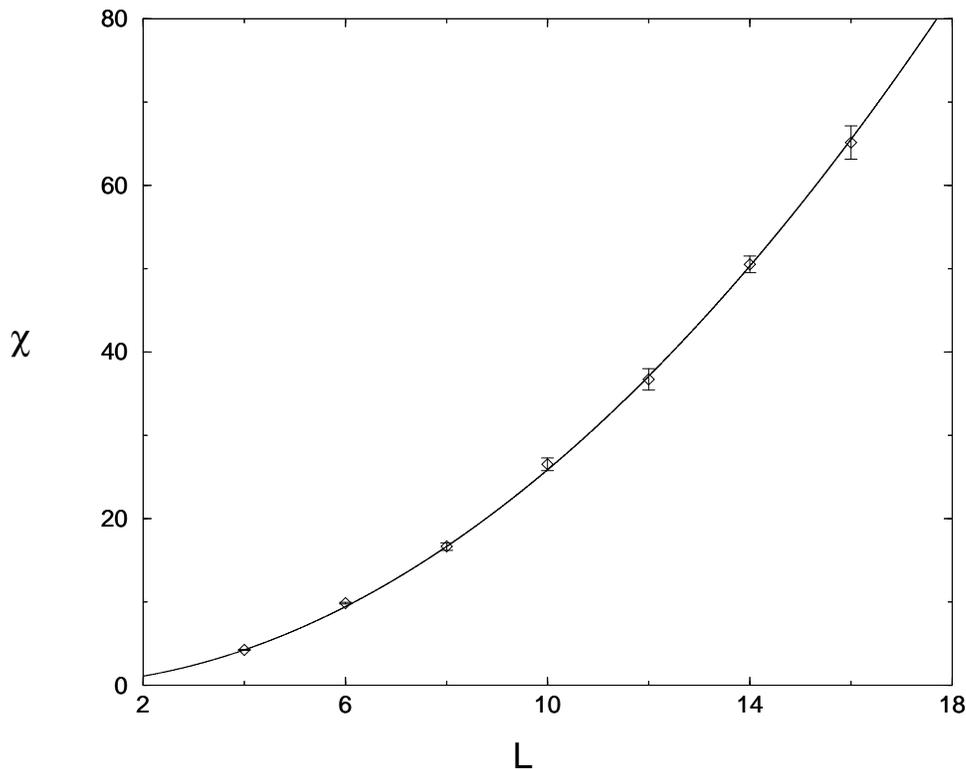,height=5in,width=4in,angle=270}
\caption{\it Finite-size scaling behavior of the chiral susceptibility $\chi$ 
as a function of the spatial size $L$ at the estimated critical inverse 
temperature $\beta_c = 0.948$. A fit of the volume-dependence (solid line) 
gives the critical exponent $\gamma/\nu = 1.98(2)$ of the 3-d Ising model.}
\end{center}
\end{figure}

\section{Conclusions}

We have applied a new fermion simulation technique --- the meron-cluster
algorithm --- to a model of staggered fermions. In contrast to standard methods
which integrate out the fermions and are left with a non-local bosonic action,
we describe the fermions in a Fock state occupation number basis and thus keep
a local bosonic action. The fermion permutation sign arises due to the 
Pauli exclusion principle as a non-local factor associated with non-trivial 
topology of the fermion world-lines. The sign leads to severe cancellations
which makes standard simulation techniques impossible to use. The decomposition
of a configuration into clusters allows us to disentangle the complicated
topology of the fermion world-lines. In particular, a meron-cluster identifies
a pair of configurations with equal weight and opposite sign. This results in
an explicit cancellation of two contributions $\pm 1$ to the path integral, 
such that only the zero-meron sector contributes to the partition function. 
Observables like the chiral susceptibility $\chi$ receive contributions only 
from the zero- and two-meron sectors. To measure $\chi$ one can hence 
eliminate all sectors with more than two merons, which leads to an exponential 
gain in statistics and to a complete solution of the fermion sign problem.

The meron-cluster algorithm allowed us to simulate a staggered fermion model 
for which standard fermion methods suffer from a severe sign problem. The model
has two flavors and a $\Z(2)$ chiral symmetry which is spontaneously broken at 
low temperatures. Applying finite-size scaling methods to the high-precision 
numerical data for the chiral susceptibility, we extracted critical exponents 
compatible with those of the 3-d Ising model. This is the expected behavior 
based on universality arguments and dimensional reduction. It would be 
interesting to apply our method to $(2+1)$ dimensions. The $N = 4$ flavor case 
was studied in \cite{Kog99} and it was verified that the model is in the 2-d 
Ising universality class. The standard fermion algorithm that was used in that 
study does not work for $N < 4$ due to the fermion sign problem. The 
meron-cluster algorithm solves the sign problem and can be used to explore 
those models.

Even in cases without a sign problem, the meron-cluster algorithm is more
efficient than standard fermion simulation methods. However, one should keep in
mind that the meron-cluster algorithm is not always applicable. For example, 
the meron concept applies only when the clusters are independent in their 
effect on the fermion sign. In addition, it must always be possible to flip the
clusters such that one reaches a reference configuration with a positive sign. 
Otherwise, some contributions from the zero-meron sector could be negative. 
For example, in our model these restrictions led to $G \geq 1$, i.e. to a 
sufficiently strong four-fermion coupling. In this paper, we have not attempted
to take the continuum limit of the lattice theory. Instead, we have studied the
chiral phase transition at a finite lattice spacing corresponding to $G = 1$. 
The restriction $G \geq 1$ would prevent us from approaching the continuum 
limit if it corresponds to $G_c < 1$. It should be pointed out that we can 
study the universal behavior of the chiral transition without taking the 
continuum limit. Of course, it is not excluded that appropriate modifications 
of the algorithm might allow us to work at $G < 1$.

A natural arena for applications of the meron-cluster algorithm are quantum
link models \cite{Cha97} which are used in the D-theory formulation of QCD 
\cite{Bro97,Bea98,Wie99}. In D-theory continuous classical fields arise from 
the dimensional reduction of discrete quantum variables. In these models the 
continuum limit is approached by varying the extent of an extra dimension, not 
by varying the value of a bare coupling constant. Hence, restrictions such as 
$G \geq 1$ would not prevent us from taking the continuum limit. In quantum 
link QCD the quarks arise as domain wall fermions. The application of 
meron-cluster algorithms to domain wall fermions is in progress. Also there are
many applications to sign problems in condensed matter physics. Investigations 
of antiferromagnets in a magnetic field and of systems in the Hubbard model 
family will be reported elsewhere.

At present, the meron-cluster algorithm is the only method that allows us to
solve the fermion sign problem. A severe sign problem arises in lattice QCD
calculations at non-zero baryon number due to a complex action. It is therefore
natural to ask if our algorithm can be applied to this case. At non-zero 
chemical potential the 2-d $O(3)$ model, which is a toy model for QCD, also 
suffers from a sign problem due to a complex action. When applied to the 
D-theory formulation of this model, the meron-cluster algorithm solves the sign
problem completely. It remains to be seen if a similar result can be achieved 
for QCD.

\section*{Acknowledgements}

S. C. likes to thank M.I.T. and U.-J. W. likes to thank Duke University for
hospitality. We have benefitted from discussions about cluster algorithms with 
R. C. Brower, H. G. Everts and M. Troyer. U.-J. W. also likes to thank the 
A. P. Sloan foundation for its support.

\end{document}